\documentclass[pdflatex,sn-mathphys]{sn-jnl}

\usepackage{aas_macros}

\jyear{2021}%

\theoremstyle{thmstyleone}%
\theoremstyle{thmstyletwo}%

\theoremstyle{thmstylethree}%
\usepackage{amsmath}
\usepackage{ragged2e}

\usepackage[table]{xcolor}
\usepackage[normalem]{ulem}

\definecolor{phelipeorange}{RGB}{230,120,20} 

\definecolor{berblue}{RGB}{20,110,220}

\begin{document}

\title[
AI Perspective into Time-domain Astronomy]{
\centering{Toward decision-aware AI for LSST-scale time-domain astronomy}
}



\author*[1,2]{\fnm{C. R.} \sur{Bom}}\email{debom@cbpf.br} 
\author[3]{\fnm{A. } \sur{Mahabal}}
\author[4,5]{\fnm{F.} \sur{Bianco}}
\author[1]{\fnm{P.} \sur{Darc}}
\author[1]{\fnm{B.} \sur{Fraga}}

\author[6]{\fnm{R.} \sur{Bonito}}
\author[4,5]{\fnm{S.} \sur{Chaini}}
\author[7]{\fnm{M. W.} \sur{Coughlin}}
\author[8]{\fnm{S.} \sur{Dillmann}}
\author[9,10]{\fnm{F.} \sur{Fontinele Nunes}}
\author[11]{\fnm{A.} \sur{Gomboc}}
\author[12]{\fnm{N.} \sur{Hernitschek}}
\author[13]{\fnm{X.} \sur{Li}}
\author[14]{\fnm{F. Z.} \sur{Majidi}}
\author[15]{\fnm{A.I.} \sur{Malz}}
\author[16]{\fnm{A.} \sur{Melandri}}
\author[14]{\fnm{V.} \sur{Petrecca}}
\author[16]{\fnm{S.} \sur{Piranomonte}}
\author[17]{\fnm{M.} \sur{Rabus}}
\author[18]{\fnm{F.} \sur{Ragosta}}
\author[19]{\fnm{O.} \sur{Razim}}
\author[20]{\fnm{M. C.} \sur{Rom\~ao}}
\author[21,22]{\fnm{N.} \sur{Sarin}}
\author[9]{\fnm{A.} \sur{Sasli}}
\author[23]{\fnm{V.A.} \sur{Sre\' ckovi\' c}}
\author[6,24]{\fnm{A.} \sur{Tramuto}}
\author[25]{\fnm{V.} \sur{Vuj\v ci\' c}}
\author[26]{\fnm{M. J.} \sur{Vyas}}
\author{\fnm{} \sur{Rubin LSST Transients and Variable Stars
Science Collaboration}}

\affil[1]{\orgname{Artificial Intelligence for Physics Laboratory, Centro Brasileiro de Pesquisas F\'isicas}, \orgaddress{\street{Rua Dr. Xavier Sigaud 150}, \city{Rio de Janeiro}, \postcode{22290-180}, \state{RJ}, \country{Brazil}}}
\affil[2]{\orgname{Laboratorio Interinstitucional de e-Astronomia-LIneA}, \orgaddress{\street{Rua Gal. José Cristino 77}, \city{Rio de Janeiro}, \state{RJ}
\postcode{20921-40}, \country{Brazil}}}

\affil[3]{\orgname{Division of Physics, Mathematics and Astronomy, California
Institute of Technology}, \orgaddress{, \city{Pasadena}, \state{CA}
\postcode{91125}, \country{USA}}}

\affil[4]{\orgname{University of Delaware, Department of Physics and Astronomy}, \orgaddress{\city{Newark}, \state{DE}, \postcode{19716}, \country{USA}}}

\affil[5]{\orgname{University of Delaware, Data Science Institute}, \orgaddress{\city{Newark}, \state{DE}, \postcode{19716}, \country{USA}}}

\affil[6]{\orgname{INAF--Osservatorio Astronomico di Palermo}, \orgaddress{\street{Piazza del Parlamento 1}, \city{Palermo}, \postcode{90134}, \country{Italy}}}

\affil[7]{\orgname{University of Minnesota}, \orgaddress{\city{Minneapolis}, \state{MN}, \postcode{55455}, \country{USA}}}

\affil[8]{\orgname{Stanford University}, \orgaddress{\street{450 Jane Stanford Way}, \city{Stanford}, \state{CA}, \postcode{94305}, \country{USA}}}

\affil[9]{\orgname{School of Physics and Astronomy, University of Minnesota}, \orgaddress{\city{Minneapolis}, \state{MN}, \postcode{55455}, \country{USA}}}

\affil[10]{\orgname{NSF Institute on Accelerated AI Algorithms for Data-Driven Discovery (A3D3)}, \orgaddress{\city{Minneapolis}, \state{MN}, \postcode{55455}, \country{USA}}}

\affil[11]{\orgname{Center for Astrophysics and Cosmology, University of Nova Gorica}, \orgaddress{\street{Vipavska 13}, \city{Nova Gorica}, \postcode{5000}, \country{Slovenia}}}

\affil[12]{\orgname{CITEVA, Universidad de Antofagasta}, \orgaddress{\street{Avenida U. de Antofagasta 02800}, \city{Antofagasta}, \postcode{1270300}, \country{Chile}}}

\affil[13]{\orgname{Johns Hopkins University}, \orgaddress{\street{3400 N. Charles St.}, \city{Baltimore}, \state{MD}, \postcode{21218}, \country{USA}}}

\affil[14]{\orgname{INAF--Osservatorio Astronomico di Capodimonte}, \orgaddress{\street{Salita Moiariello 16}, \city{Napoli}, \postcode{80131}, \country{Italy}}}

\affil[15]{\orgname{Space Telescope Science Institute}, 
\orgaddress{\street{3700 San Martin Drive}, \city{Baltimore}, \state{MD}, \country{USA}}}

\affil[16]{\orgname{INAF--Osservatorio Astronomico di Roma}, \orgaddress{\street{Via di Frascati 33}, \city{Monte Porzio Catone}, \postcode{00078}, \country{Italy}}}

\affil[17]{\orgname{Departamento de Matem{\'a}tica y F{\'i}sica Aplicadas, Facultad de Ingenier{\'i}a, Universidad Cat{\'o}lica de la Sant{\'i}sima Concepci{\'o}n}, \orgaddress{\street{Alonso de Rivera 2850}, \city{Concepci{\'o}n}, \country{Chile}}}

\affil[18]{\orgname{Dipartimento di Fisica ``Ettore Pancini'', Universit\`a di Napoli Federico II}, \orgaddress{\street{Via Cinthia 9}, \city{Napoli}, \postcode{80126}, \country{Italy}}}

\affil[19]{\orgname{University of Nova Gorica}, \orgaddress{\street{Vipavska 13}, \city{Nova Gorica}, \postcode{5000}, \country{Slovenia}}}

\affil[20]{\orgname{Institute for Particle Physics Phenomenology, Durham University}, \orgaddress{\city{Durham}, \postcode{DH1 3LE}, \country{UK}}}

\affil[21]{\orgname{Kavli Institute for Cosmology, University of Cambridge}, \orgaddress{\street{Madingley Road}, \postcode{CB3 0HA}, \country{UK}}}

\affil[22]{\orgname{Institute of Astronomy, University of Cambridge}, \orgaddress{\street{Madingley Road}, \postcode{CB3 0HA}, \country{UK}}}

\affil[23]{\orgname{Institute of Physics Belgrade}, \orgaddress{\street{Pregrevica 118}, \city{Belgrade}, \country{Serbia}}}

\affil[24]{\orgname{Dipartimento di Fisica e Chimica ``E. Segr\`e'', Universit\`a degli Studi di Palermo}, \orgaddress{\street{P.zza del Parlamento 1}, \city{Palermo}, \postcode{90134}, \country{Italy}}}

\affil[25]{\orgname{Astronomical Observatory Belgrade}, \orgaddress{\street{Volgina 7}, \city{Belgrade}, \postcode{11060}, \country{Serbia}}}

\affil[26]{\orgname{Department of Astrophysics, University of Vienna}, \orgaddress{\street{T\"urkenschanzstra\ss e~17}, \city{Vienna}, \postcode{1180}, \country{Austria}}}





\abstract{

The Vera C. Rubin Observatory’s Legacy Survey of Space and Time (LSST) will generate approximately \(10^7\) alerts per night, pushing time-domain astronomy beyond pipelines that treat discovery as a static labeling problem. We argue that LSST is better understood as a partially observed dynamical environment, in which scientific return depends on the quality of follow-up decisions made under uncertainty and finite observational resources. The central challenge is therefore to maintain evolving, uncertainty-aware representations of astrophysical sources and to select actions that maximize long-term scientific value. We propose that foundation models trained on heterogeneous time-domain data can learn survey-scale representations of source state, while decision-theoretic policies support principled, auditable allocation of follow-up resources. Embedded within human-supervised agentic systems, these components position AI as part of the operational inference loop rather than as a downstream predictive tool. The way such systems represent belief, optimize utility, and expose their reasoning will shape observational efficiency, the distribution of scientific agency, including who participates in discovery and the scientific questions that receive priority.
}

\keywords{AI, Time-domain Astronomy, Photometry, Light Curve Analysis, Spectroscopic Follow-up, Classification}

\maketitle

\section{Introduction}

\label{sec1}
 
Time-domain astronomy has historically operated under conditions of scarcity. Transient events were rare, sky coverage limited, and cadences modest. The past decade marked a transitional phase: surveys such as ZTF increased alert rates to $10^5-10^6$ per night while maintaining latency low enough to enable systematic follow-up for selected subsets \citep{Bellm2019,Masci2019,Patterson2019}. Machine learning classifiers became operationally indispensable, yet human oversight remained feasible for curated alert streams \citep{Masci2019}.
LSST crosses a qualitative threshold. Generating approximately $10^7$ alerts per night spanning greater depth, colour space (from multi-band filters) and phenomenological diversity \citep{Ivezic2019,Graham2024DMTN102}, manual intervention becomes infeasible across most stages of discovery. Early light curves will often consist of only a handful of sparse, irregular observations, spectroscopic ground truth will be available for a small fraction of events, and survey strategy and population demographics will evolve over a decade \citep{Ivezic2019, lsst_cadence}.

These conditions expose a problem deeper than scale alone. Consider an early transient detected in two bands over 24 hours, consistent with a young Type Ia supernova, a core-collapse event, a tidal disruption event or an unusual AGN flare. Its scientific value depends less on its most probable class than on whether timely follow-up can distinguish among these interpretations before diagnostic information is lost. At LSST scale, such cases are routine. LSST is therefore better understood as a partially observed environment in which alerts update knowledge of evolving sources and follow-up actions determine what can be learned next.

\section{The limits of supervised classification-centric machine learning}\label{sec2}

Machine learning is already deeply embedded in time-domain astronomy. Classifiers such as RAPID \cite{Muthukrishna2019} and ParSNIP \cite{Boone2021}, SuperNNova\cite{moller2020}   and CATS \citep{fraga_cats} have demonstrated strong performance on benchmark datasets, and alert brokers including ALeRCE \citep{Foster2021}, ANTARES \citep{antares} and Fink \citep{Moller2021} have built classification pipelines into operational infrastructure.  At LSST scale, however, they reveal a structural limitation: current approaches are optimized for static prediction tasks, whereas the scientific problem is dynamic and decision-dependent.

The prevailing paradigm treats each alert as an independent classification problem, evaluated through accuracy, purity or area under the Receiver Operating Characteristic (ROC) curve \citep[e.g.][]{villar2019, fraga_cats, moller2020, 2018ApJS..236....9N}. These metrics implicitly equate scientific value with correct labeling and treat alerts in isolation. Neither assumption holds. For the early transient above, a $60\%$ Type Ia probability does not determine whether spectroscopy should be triggered before the diagnostic phase is lost or deferred for another target. The cost of a wrong action is asymmetric: missing early colour evolution in a rare core-collapse supernova is irreversible, whereas delaying spectroscopy of a well-understood Type Ia may carry little scientific penalty. Per-alert accuracy encodes none of this structure.

More fundamentally, alerts are not independent. Follow-up decisions shape the future information available to the survey and influence population-level inference. 
The relevant objective is expected scientific utility: what action would most improve knowledge, given the current uncertainty about the object?

\section{Foundation Models as Scientific Priors}

Rather than training task-specific classifiers tied to predefined taxonomies, foundation models learn general representations of variability across heterogeneous datasets. Trained in largely self-supervised fashion on diverse light curves and contextual information \citep[e.g.;][]{found01}, such models encode statistical regularities that function as survey-scale scientific priors. Early efforts, including AstroCLIP \citep{Parker2024} and AstroPT \citep{smith2024astropt}, demonstrate that meaningful latent structure can be recovered from photometric data alone, without relying on labelled training sets that will always be incomplete at LSST scale.

The latent state of a source extends beyond class membership to include evolutionary phase, characteristic timescales, colour evolution, luminosity history and associated uncertainties.
A foundation model seeks to embed an early transient within a representation space learned from large populations of historical variability, encoding proximity to multiple evolutionary pathways, uncertainty along physically meaningful directions, and the degree of support from prior data. Such information could be more actionable than a single class probability, because it informs the decision system about what is known, what remains uncertain, and which additional observations are likely to be most informative.

The nuclear transient case illustrates a complementary strength. A foundation model trained on long-baseline variability can encode whether a galactic flare is consistent with the source's own historical structure function, whether it departs from typical stochastic AGN behaviour at a statistically meaningful level, and whether the departure is accelerating or transient \citep{2021ApJ...908....4V}. This longitudinal context, which a static classifier discards by design, may be the most diagnostic information available in the absence of spectroscopy. 

More generally, a shared latent space can support multiple downstream objectives without task-specific retraining \citep{Bengio2013,Parker2024}; in astronomy, such representations can be coupled to active anomaly discovery and prioritization pipelines \citep{astromaly}, while domain-adaptation methods help mitigate brittleness under changing survey conditions \citep{Pandya_2025}.

To guide follow-up, these representations must be embedded in temporal or causal models able to forecast what alternative observations would reveal.
Agentic systems \citep{Gridach2025} can then evaluate candidate actions \textit{in silico}, using simulators and learned causal or world models \citep{richens2025general,ding2025understanding} to compare plausible observing strategies before committing scarce observational resources. 
Incorporating metadata during training could further improve robustness to temporal and instrumental variation, and may help the model track shifts in observing conditions over time.

Such approaches support counterfactual reasoning: given the current belief state for a source, what would spectroscopy at this epoch reveal relative to additional photometry tomorrow? The ability to simulate the informational consequences of alternative actions turns a representation model into a substrate for decision-making. Deploying such models in LSST operations will require accommodating non-stationarity, continual learning and robustness to distribution shift without catastrophic forgetting \citep{Kirkpatrick2017,VanDeVen2024}; furthermore, for transient phenomena, timing is crucial. Such systems would benefit the most by acting as soon as possible in the alert stream.

\section{Active learning and agentic orchestration for decision-aware discovery}

At LSST scale, learning does not end at deployment. Follow-up observations including spectroscopy, rapid multi-band photometry, cadence changes and target-of-opportunity triggers, are interventions that reshape the information available for inference. Each is costly, irreversible and informative. Active learning therefore becomes an operational principle: the aim is to select the observations that yield the greatest scientific gain under finite resources, particularly for early-time, ambiguous or unusual events~\citep{2020arXiv201005941K, 2025PASA...42...57M}.

\begin{figure*}[t]
    \centering
    \includegraphics[width=1\linewidth]{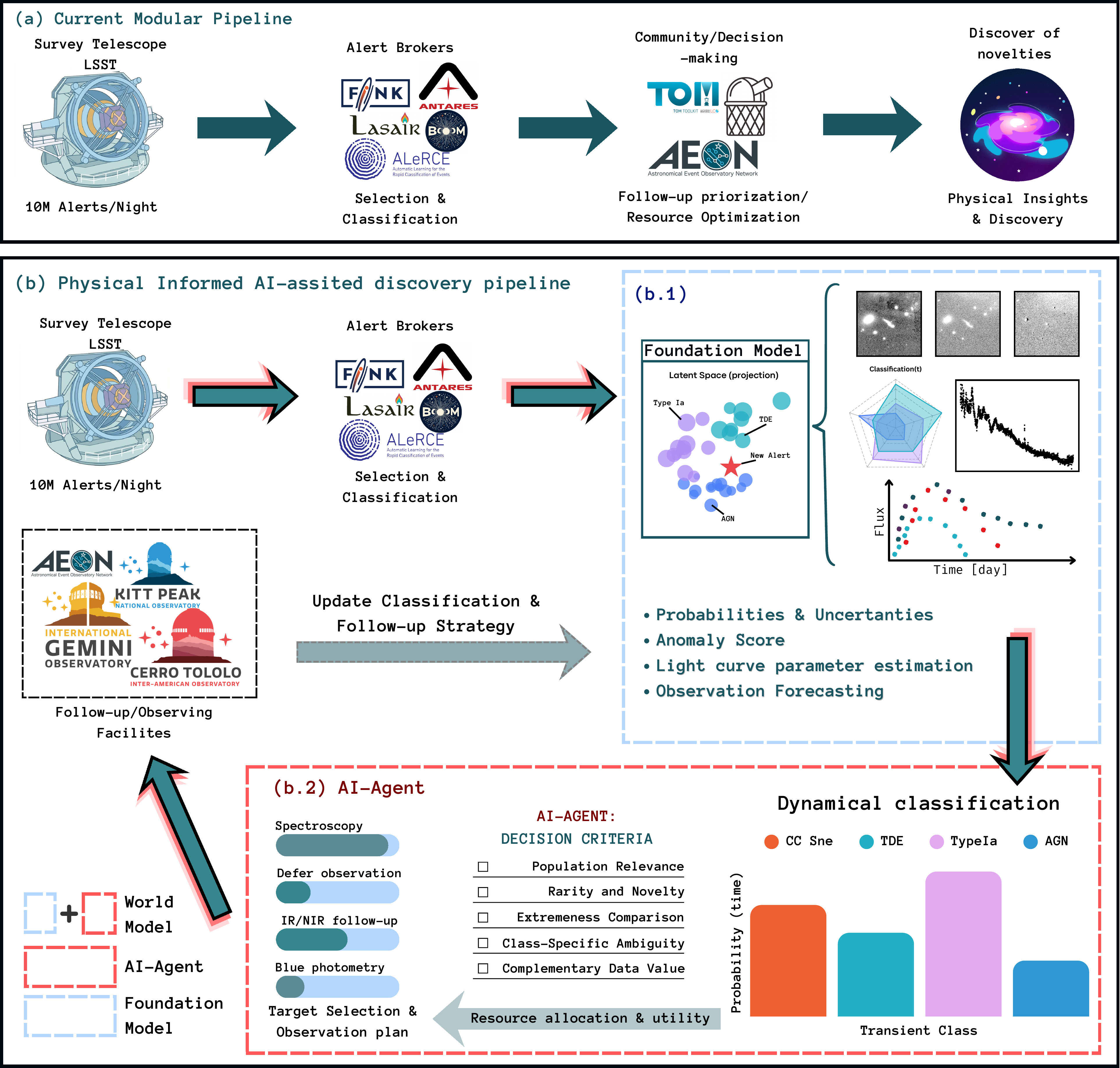}
    
    \caption{\textbf{From modular alert handling to an integrated AI discovery pipeline. a}, Current workflow: LSST alerts pass through brokers for selection and classification before community target-management and scheduling systems prioritize follow-up, with limited propagation of a shared probabilistic state. \textbf{b}, Proposed decision-aware workflow: a foundation model maps multimodal survey data into a latent source-state representation, including probabilities, uncertainties, anomaly scores, light-curve parameters and observation forecasts. An AI agent combines these outputs with explicit decision criteria and utility-based resource allocation to recommend spectroscopy, blue photometry, infrared follow-up or deferral. New observations update both classification and strategy. The foundation model and agent could also be integrated as a world model coupling representation, inference and action.}

    \label{fig:pipeline}
\end{figure*}

A classical Partially Observed Markov Decision Process (hereafter POMDP) provides a useful starting point for LSST alert handling, which requires acting under uncertainty, updating a belief state as new data arrive, and selecting interventions sequentially \citep{kaelbling1998,sutton1998}. In this setting, the belief state may summarize latent source properties, competing physical hypotheses, associated uncertainties, temporal evolution and, where available, host-galaxy context \citep{Shah2025}. Yet this abstraction does not fully capture the context in which scientific choices are made. Decisions about whether to follow a transient depend on uncertainty in its inferred physical state and on broader epistemic factors: whether the event addresses an open question, whether similar systems are already well represented in the literature, and how strategically informative a new observation would be. Such considerations are dynamic and difficult to encode in a fixed state or reward function. Agentic orchestration may therefore offer a more faithful abstraction for LSST-scale discovery, combining probabilistic inference with retrieval, memory and context-sensitive utility evaluation to select among actions such as spectroscopic follow-up, additional photometric epochs, broker-level prioritization and deliberate deferral. In this view, inference and intervention become part of a single decision loop.
Figure \ref{fig:pipeline}  contrasts the current broker-centred workflow with a decision-aware inference–action loop coupling foundation-model representations, agentic inference, utility-based policy selection and follow-up feedback

Each action carries a cost and yields an observation that updates the system’s state of knowledge through Bayesian inference. For the 24-hour transient described above, the system must decide whether spectroscopy justifies displacing another target on an 8-metre facility, whether a bluer photometric epoch would provide sufficient information at lower cost, or whether deferral is acceptable. These trade-offs require estimating the informational value of candidate actions under an explicit utility function, conditioned on the inferred source state and broader scientific context \citep{foster2021deep}.

\begin{figure}
    \centering
    \includegraphics[width=1\linewidth]{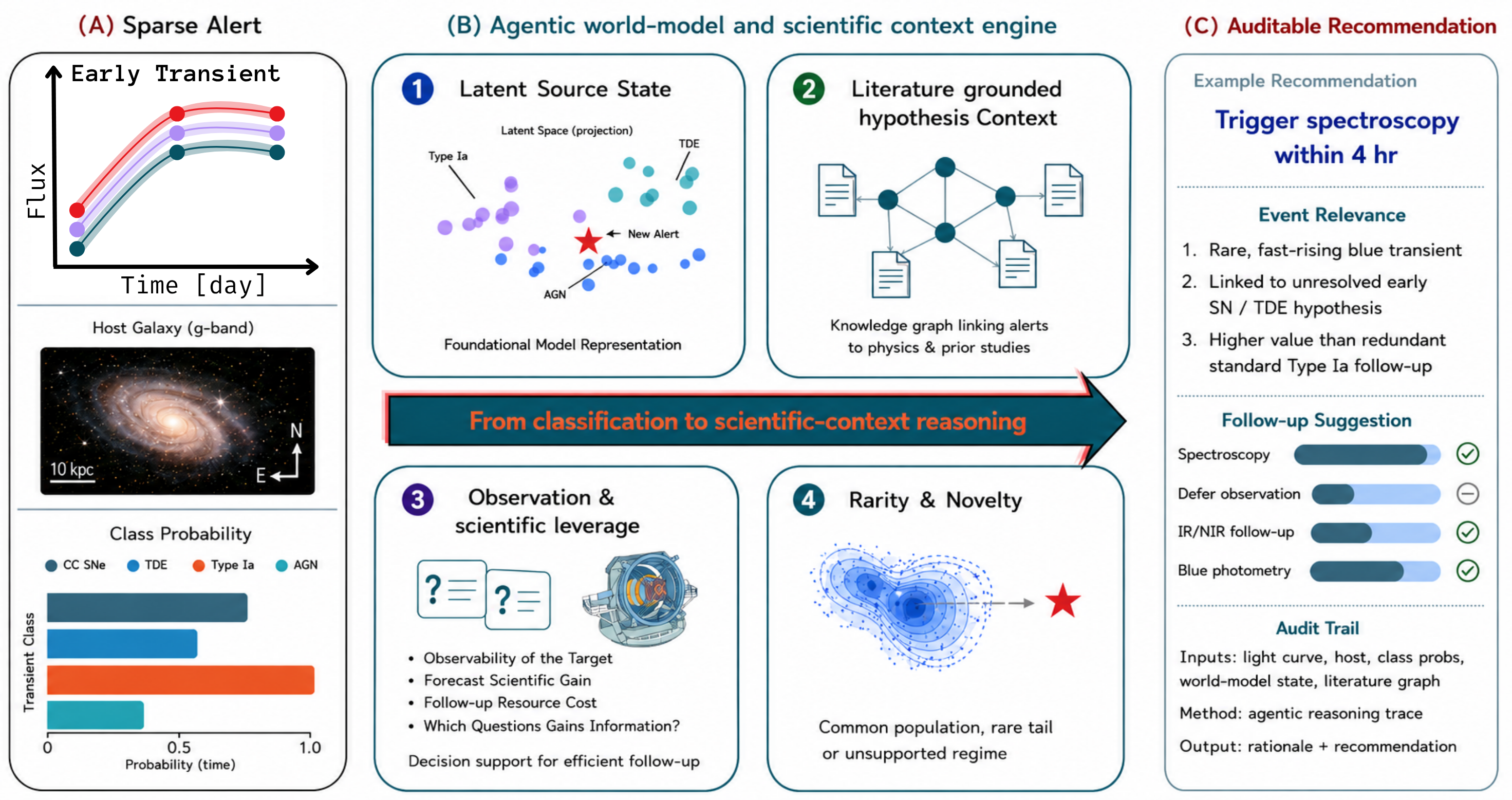}
    
     \caption{\textbf{From classification to scientific-context reasoning}. A sparse alert with limited early photometry, host context and ambiguous class probabilities is evaluated by an agentic model and scientific-context engine. The system embeds the event in a latent source-state representation, compares it with prior populations and unsupported regimes, retrieves literature-grounded hypotheses, estimates scientific leverage and forecasts possible follow-up actions. The output is an auditable recommendation rather than a class label alone; here, spectroscopy is prioritized because the event is rare, time-sensitive and informative for distinguishing supernova, tidal-disruption-event or AGN interpretations.}

    \label{fig:reasoning}
\end{figure}

Fully optimal decision-making is unlikely to be tractable at LSST scale. Practical systems will therefore rely on approximations: myopic information-gain estimates, hierarchical triage that reserves expensive optimization for high-value candidates, and learned policies trained in simulation under realistic cadence and follow-up constraints \citep{Foster2021}. Figure \ref{fig:reasoning} illustrates this shift for an individual event: the output is not a class label alone, but an auditable recommendation combining latent-state placement, host context, rarity, literature-grounded hypotheses, expected information gain and follow-up time sensitivity.

Some degree of agency is therefore operationally necessary. Existing brokers and target-management systems already contain pieces of this capability; the missing step is integration: linking perception, memory, inference and action through explicit, auditable utility functions.
Such a system need not be autonomous in an unrestricted sense. Its autonomy can remain bounded by predefined resource limits, human-specified scientific goals and oversight mechanisms designed to keep system behaviour aligned with community intent \citep{hadfield2016cooperative}. In this formulation, researchers define the scientific questions and strategic priorities, the agentic system evaluates candidate interventions against those goals, and actions are proposed subject to human-defined constraints. Agency, in this sense, is a tool for preserving scientific coherence under scale rather than a licence for unconstrained decision-making.

\section{Implications for AI-assisted astronomy}

This shift will be especially consequential in multi-messenger and multi-facility settings. Rubin will operate alongside gravitational-wave detectors, electromagnetic follow-up facilities and, increasingly, neutrino and gamma-ray observatories \citep{andreoni2022target,nicholl2025electromagnetic,bom2024designing}. Events such as neutron-star mergers or extreme nuclear transients may generate alerts with different latencies, localization uncertainties and follow-up requirements. Coordinated response within the relevant time windows will increasingly require decision-aware systems that maintain shared probabilistic context across messenger channels and facilities \citep{almualla2020dynamic}.

The transition does not imply uniform automation. Anomalies, physically ambiguous events and judgments of broader significance should remain human-led, because scientific relevance is context-dependent, difficult to specify exhaustively and often depends on the human capacity for out-of-domain generalization. In astronomy, anomaly detection systems already show that expert feedback remains essential for distinguishing scientifically interesting outliers from artefacts or merely rare but uninformative events \citep{astromaly,fluke2020understanding}. By contrast, rapid response, population-level triage and multi-facility queue coordination are likely to rely increasingly on bounded autonomy. The design challenge is therefore to build interfaces that keep humans meaningfully in the loop at the supervisory and high-level interpretive layers, while delegating execution to agentic systems capable of operating at survey scale.

Similar patterns are emerging across data-intensive science: real-time event selection at the LHC, active learning in biological sequence design, and decision-theoretic sensor placement in environmental monitoring all share a common structure: partial observability, costly or irreversible interventions, and tightly coupled inference and action \citep{albertsson2018machine,pandi2022versatile,krause2008near}. Methods developed for LSST are therefore likely to be relevant well beyond astronomy, just as advances from other fields should be imported more systematically into astronomical practice. More broadly, recent successes in AI for science highlight the value of cross-domain methodological transfer, even when the underlying scientific questions differ \citep{baker2019workshop}. The design of AI-assisted discovery systems is itself a scientific problem: choices about utility functions, uncertainty representation and the boundary between AI and human decision-making will shape what gets discovered, what gets missed and what questions the community learns to ask.

\section{Governance, centralization and scientific formation}

\par Decision-system architecture will shape not only efficiency, but participation, priorities and scientific influence. 
Foundation models and stateful agents demand substantial compute, engineering capacity and supporting infrastructure, none of which is evenly distributed. If these systems become core scientific dependencies, influence over their design may concentrate within a small number of well-resourced institutions, with downstream consequences for who can meaningfully participate in discovery \citep{bommasani2021opportunities}.

The concern is epistemic as well as institutional: decision systems encode judgments about what is worth detecting, ranking and pursuing.
Philosophy of science has long argued that scientific reasoning is not fully separable from value judgments, particularly under uncertainty \citep{Douglas2000-DOUIRA}. Nor is the concentration of agenda-setting within particular groups new. Sociology of science has shown that scientific fields may be shaped by unequal scientific capital, paradigms, community norms and asymmetries in epistemic authority, which together influence which questions, methods and, increasingly, algorithms are treated as legitimate \citep{Bourdieu2004}.

Autonomous agents may amplify this dynamic by hard-coding triage criteria into follow-up pipelines. Under conditions of scarce follow-up resources, such systems determine which alerts are pursued, deferred, or missed. In time-domain astronomy, pipelines optimized for specific science cases, such as supernova cosmology, will naturally favour events most relevant to that objective, whereas unusual transients can remain under-prioritized unless anomaly-detection modules are explicitly included \citep{DJORGOVSKI201695,2023arXiv231117143H,Ishida2021}. The contribution of diverse research groups with different science cases can alleviate some of these biases.

Utility functions, training data and policy objectives should be treated as community-facing design choices subject to open scrutiny. Citizen-science programmes and more inclusive forms of governance can help broaden participation beyond the best-resourced institutions and countries \citep{marshall2015ideas,paleco2021inclusiveness,wehn2024opening}.

A subtler concern involves scientific formation. Astronomers have historically developed intuition about uncertainty by navigating it directly, triaging ambiguous events, deciding when to trigger expensive follow-up, learning from the consequences \citep{collins2019tacit}. This hands-on engagement is how the community develops the tacit knowledge required to recognize when a system is behaving unexpectedly. If decision processes become abstracted behind automated outputs, early-career researchers may be less equipped to identify failure modes or reason effectively in regimes where those systems do not apply \citep{parasuraman2010complacency,gil2017thoughtful}. Interfaces that expose belief states and utility trade-offs, simulation environments for developing decision intuition, and training programmes addressing the reasoning underlying automated policies are practical responses, not to slow automation, but to ensure the tacit knowledge required to govern these systems continues to be cultivated.

These considerations converge on non-negotiable design principles: decision logs recording which actions were taken against which objectives; interpretable utility functions that can be inspected and contested; mechanisms for human override at multiple system levels; and governance structures distributing participation in objective definition beyond well-resourced institutions and countries. At LSST scale, these are not optional enhancements but are preconditions for ensuring that scale amplifies rather than constrains the scope of astronomical discovery.

\section{Operational implications and next steps}

We outline a progressive pathway from immediate steps to longer-term architectural goals.
The most immediately actionable step is for scientific collaborations to articulate explicit decision objectives. In cosmological supernova analysis, utility may correspond to expected reduction in posterior uncertainty on the dark energy equation-of-state. In a rare transient search, it may scale with the probability that follow-up distinguishes an out-of-distribution event before its diagnostic phases are lost. 
Making such objectives explicit allows competing scientific goals to be weighed transparently and provides a basis for comparing alternative follow-up policies. Evaluation should therefore extend to decision-level performance, including cumulative scientific return, efficiency relative to benchmark strategies, and recovery of population-level parameters, while also recognizing that some scientifically important cases are difficult to encode in formal objectives and may otherwise be underrepresented.

Simulation infrastructure for this kind of benchmarking is already emerging. Rubin’s survey-strategy evaluation framework \citep{jones2014lsst}, together with the OpenUniverse2024 simulation suite \citep{openuniverse2025openuniverse2024}, provides a realistic basis for testing decision policies across both Rubin and Roman. OpenUniverse2024, in particular, delivers matched simulations on a common sky, including overlapping LSST deep-drilling and Roman time-domain survey realizations with updated transient models. Broader end-to-end LSST simulation efforts, such as DESC DC2 \citep{lsst2021lsst}, remain valuable for evaluating individual pipeline components in controlled settings.

In the near term, time-domain data-analysis systems such as alert brokers could evolve toward stateful inference layers in science cases where sequential decision-making is central. \citep{nordin2019transient}, or be directly connected to agentic platforms built on foundation or world models. As shown in Fig. \ref{fig:pipeline}, brokers would remain central to alert filtering and classification, but their outputs would feed systems that track sources over time and guide follow-up decisions, rather than ending in static rankings.
Given their cost and complexity, such science-driven platforms may be shared across collaborations, provided they implement transparent policies, remain highly flexible and adaptable, and adhere to FAIR principles (Findability, Accessibility, Interoperability, and Reuse). They should expose uncertainty, temporal context, proximity to underexplored parameter space and expected information gain alongside class probabilities. Longer term, the combination of LSST and Roman is particularly compelling: Roman’s wide-field near-infrared surveys can overlap substantially with LSST sky coverage \citep{eifler2021cosmology}, and joint Rubin--Roman simulations already highlight the value of combining realistic optical and infrared information with updated transient models \citep{openuniverse2025openuniverse2024}. A decision system that maintains a joint belief state across both surveys could therefore extract substantially greater scientific return than independent pipelines. Similar arguments apply to coordination with pathfinders such as SKA and space-based high-energy monitors, where the scientific return increasingly depends on integrated, cross-facility follow-up and information-sharing infrastructures \citep{andreoni2022target,burns2023gamma,sambruna2022nasa}.

Governance and transparency should be embedded from the outset. Decision logs, interpretable policy components, and human override mechanisms should be treated as core infrastructure from the earliest stages of development. These elements are preconditions for the community trust required for the operational use of decision-aware systems.
\section{Conclusion}
LSST inaugurates a regime in which astronomical discovery unfolds through continuous inference and intervention under uncertainty. The early transient threaded through this paper (detected in two bands over 24 hours, physically ambiguous, scientifically valuable only if acted upon at the right moment) is not an edge case but a representative unit of LSST science, multiplied across ten million alerts every night for a decade. Classification alone cannot answer the question of how to respond to it. What is required is a system that maintains an evolving belief state over physical properties, evaluates the informational consequences of possible actions against explicitly articulated objectives, and proposes interventions at latencies and volumes that exceed human coordination capacity, while remaining transparent, auditable and subject to meaningful oversight.

This goal is achievable through the evolution of already existing data infrastructure. Foundation models provide the perceptual substrate. World models support counterfactual reasoning. Decision-theoretic policies translate belief states into principled follow-up recommendations. Governance structures ensure that embedded objectives reflect community priorities rather than accumulated design defaults. The pathway is iterative, and the simulation infrastructure to benchmark progress already exists.

The architectural decisions made in the coming years will shape the epistemic character of LSST science in ways that are difficult to reverse. A well-designed decision-aware infrastructure that is transparent in its objectives, inclusive in its governance, deliberate in preserving human engagement with uncertainty, could transform LSST from a data firehose into a genuinely intelligent observatory: one that learns from its own discoveries, allocates attention where it matters most and remains open to the unexpected. The history of astronomy suggests that the most important discoveries are rarely the ones that were anticipated. The systems built to process LSST data should be designed not only to find what we are looking for, but to notice what we are not.

\backmatter

\bmhead{Acknowledgments}
C.R. Bom acknowledges the financial support from CNPq (316072/2021-4) and from
FAPERJ (grants 201.456/2022 and 210.330/2022) and
the FINEP contract 01.22.0505.00 (ref. 1891/22) and 01.25.0215.00 (ref. 1021/24).
R. Bonito, and A. Tramuto acknowledges the support of the INAF grant 1.05.23.04.02 (“PACE”, PI: Bonito). MCR is supported by the STFC under Grant No. ST/T001011/1. SC acknowledges support received from the NASA FINESST program, grant 80NSSC25K0312.
This publication is co- funded by
the European Union’s Horizon Europe research and innovation program under the Marie Sklodowska-Curie COFUND Postdoctoral Programme grant agreement No.101081355- SMASH and by the Republic of Slovenia and the European Union from the European Regional Development Fund.
Co - funded by the European Union. Views and opinions
expressed are however those of the author(s) only and do not necessarily reflect those of the European Union or European Research Exacutive Agency. Neither the European Union nor the granting authority can be held responsible for them.
F.Z. Majidi acknowledges support from the ASI-INAF Agreement no. 2021-12-HH.0 and Addendum 2021-12-HH.1-2024 "Missione Solar-C EUVST-Supporto scientifico di Fase B/C/D".




\bibliography{sn-bibliography}

\end{document}